\documentclass{PoS}
\usepackage{graphicx}
\usepackage{subfig}
\usepackage{floatrow}
\usepackage{blindtext}
\usepackage{float}
\restylefloat{table}
\usepackage{lineno}
\usepackage{lipsum}
\let\OLDthebibliography\thebibliography
\renewcommand\thebibliography[1]{
  \OLDthebibliography{#1}
  \setlength{\parskip}{0pt}
  \setlength{\itemsep}{0pt plus 0.3ex}
}

\title{Search for Astrophysical Tau Neutrinos with IceCube}

\ShortTitle{Search for Astrophysical Tau Neutrinos with IceCube}

\author{\speaker{Donglian Xu} for the IceCube Collaboration%
\thanks{http://icecube.wisc.edu/}\\
Wisconsin Particle Astrophysics Center, 222 West Washington Ave., Suite 500
Madison, WI 53703, USA \\  Department of
  Physics, University of Wisconsin-Madison, 1150 University Avenue
Madison, WI  53706, USA\\
  E-mail: \email{dxu@icecube.wisc.edu}}

\abstract{High-energy (TeV-PeV) cosmic neutrinos are expected to be
  produced in extremely energetic astrophysical sources such as active
  galactic nuclei. 
The IceCube Neutrino Observatory at the South Pole has recently
detected a diffuse astrophysical neutrino flux. While the flux is
consistent with all flavors of neutrinos being present, identification
of tau neutrinos within the flux is yet to occur.  
Although tau neutrino production is thought to be low at the source, an equal fraction of neutrinos are expected at Earth due to averaged neutrino oscillations over astronomical distances. Above a few hundred TeV, tau neutrinos become resolvable in IceCube with negligible background from cosmic-ray induced atmospheric neutrinos. Identification of tau neutrinos within the observed flux is crucial to precise measurement of its flavor content, which could serve to test fundamental neutrino properties over extremely long baselines, and possibly shed light on new physics beyond the Standard Model. We present the analysis method and results from a recent search for astrophysical tau neutrinos in three years of IceCube data.}

\FullConference{38th International Conference on High Energy Physics\\
  3-10 August 2016\\
  Chicago, USA}

\begin{document}


\section{Introduction}

The IceCube Neutrino Observatory has detected a diffuse astrophysical
neutrino flux at $>$ 6$\sigma$ significance over the atmospheric
background \cite{Aartsen:2013jdh, Aartsen:2014gkd, Kopper:2015vzf}. 
To date, no neutrino point sources have been identified to
have contributed substantially to the observed flux. Assuming a
neutrino flavor ratio of $\nu_e$ : $\nu_{\mu}$ : $\nu_{\tau}$ = 1 : 2
: 0, based on pion production at the source, equal fractions of
neutrinos are expected at Earth due to standard neutrino oscillations
over astronomical distances. Other scenarios of neutrino flavor ratios
at the source ranging from 1 : 0 : 0 and 0 : 1 : 0 also all predict appreciable
numbers of tau neutrinos at Earth. Analyses combining various event
samples and performing global maximum likelihood
fits found the neutrino flavor ratio to be consistent with 1~:~1~:~1, but
with large uncertainties \cite{PhysRevLett.114.171102, Aartsen:2015knd}. The largest uncertainty resides
in the degeneracy of $\nu_e$ and $\nu_{\tau}$ events in IceCube. Identification of
$\nu_{\tau}$ will help break this degeneracy. A multitude of new physics models predict significant
deviation from equal mixing of astrophysical neutrinos
\cite{PhysRevLett.115.161303, PhysRevLett.115.161302}. A precise
measurement of the astrophysical neutrino flavor contents can help decode
the neutrino production mechanisms at source, test standard neutrino
oscillations over extremely long baselines, and constrain new physics
models beyond the Standard Model.

IceCube is a cubic kilometer neutrino observatory located at the
geographic South Pole.
It was built to detect TeV-PeV astrophysical neutrinos. The construction of IceCube
started in 2004 and was completed in December 2010. 
The IceCube detector consists of 86 cables, called {\it strings}, deployed at depths between 1450 m and 2450 m.
Each string is instrumented with 60 digital optical modules (DOMs).
This array of 5160 DOMs encompasses $\sim$1 gigaton volume in ultra transparent glacial ice, making it
the world's largest neutrino detector to date. The inter-string
distance is $\sim$125~m, and the vertical distance between two DOM is
$\sim$17~m. At the bottom center of IceCube, there is a denser
sub-array called DeepCore with an inter-string distance of $\sim$60-70~m,
and an inter-DOM distance of $\sim$7~m. DeepCore lowers the detection
energy threshold of IceCube to $\sim$10 ~GeV, enabling neutrino
oscillation physics with atmospheric neutrinos and probes of new physics such as
dark matter searches at these energies. The basic building blocks of IceCube are 
DOMs, which each include a 10 inch PMT which is housed in a high-pressure glass vessel that can withstand
pressures up to 10,000 psi. Each DOM also has processing and digitization
electronics. IceCube waveform digitization occurs in ice. There are
two types of waveform digitizers: one
called the Analog Transient Waveform Digitizer (ATWD) and the other called
the fast Analog to Digital Converter (fADC). An ATWD has three channels with
gains of (16, 2, 0.25) times the nominal gain of 10$^7$. During
digitization, the highest gain channel is initiated first to capture
the best details of the signal waveforms. The lower gain channel will be
initiated if the higher gain channels are saturated.
The ATWD digitizes at 3.3~ns per sample for 128 samples in one
waveform (422.4~ns) \cite{achterberg2006first}.  

\begin{figure*} [t]
\centering
 \subfloat{{\includegraphics[width=0.47\textwidth]{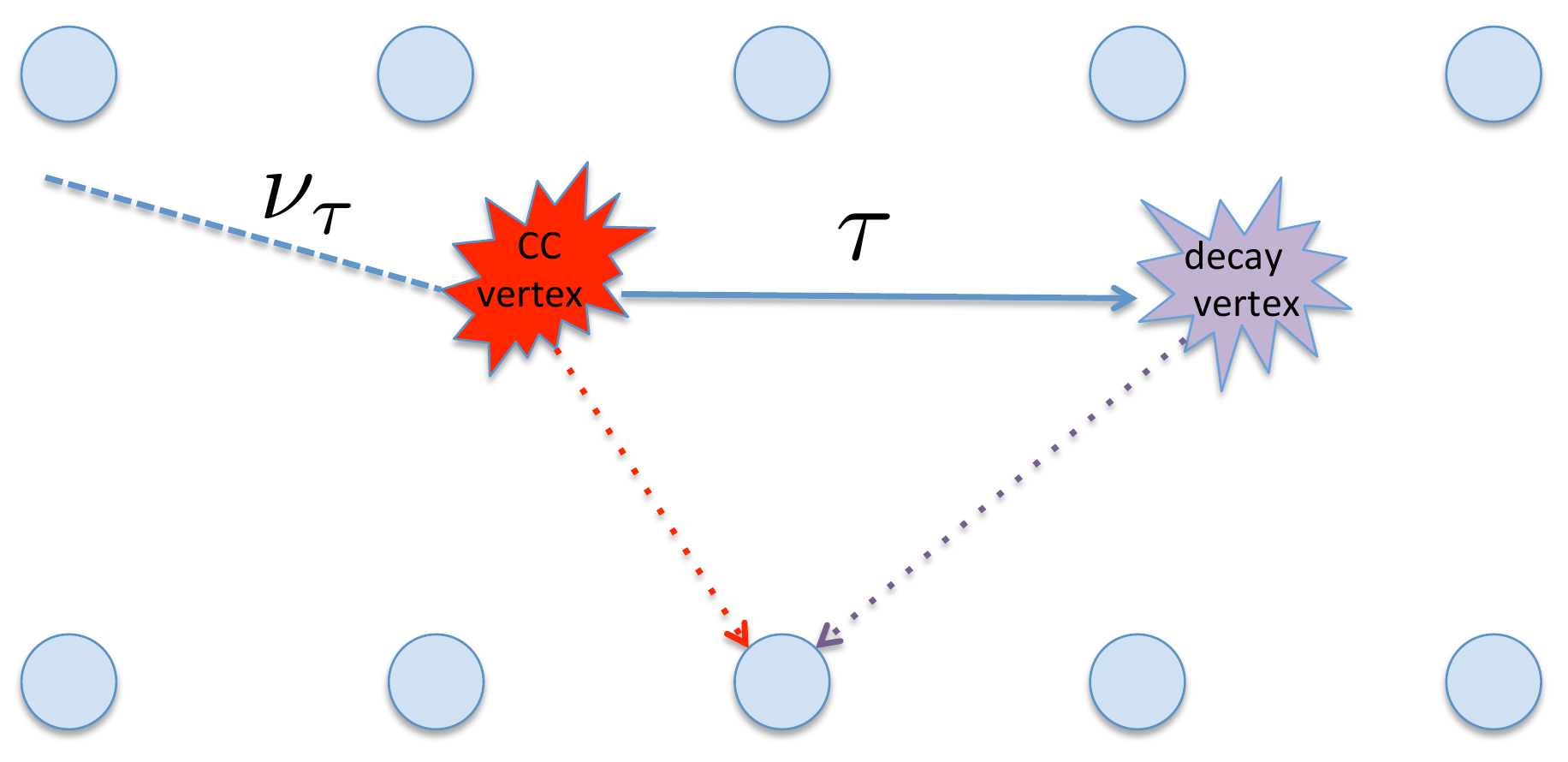}}}%
    \qquad
    \subfloat{{\includegraphics[width=0.47\textwidth]{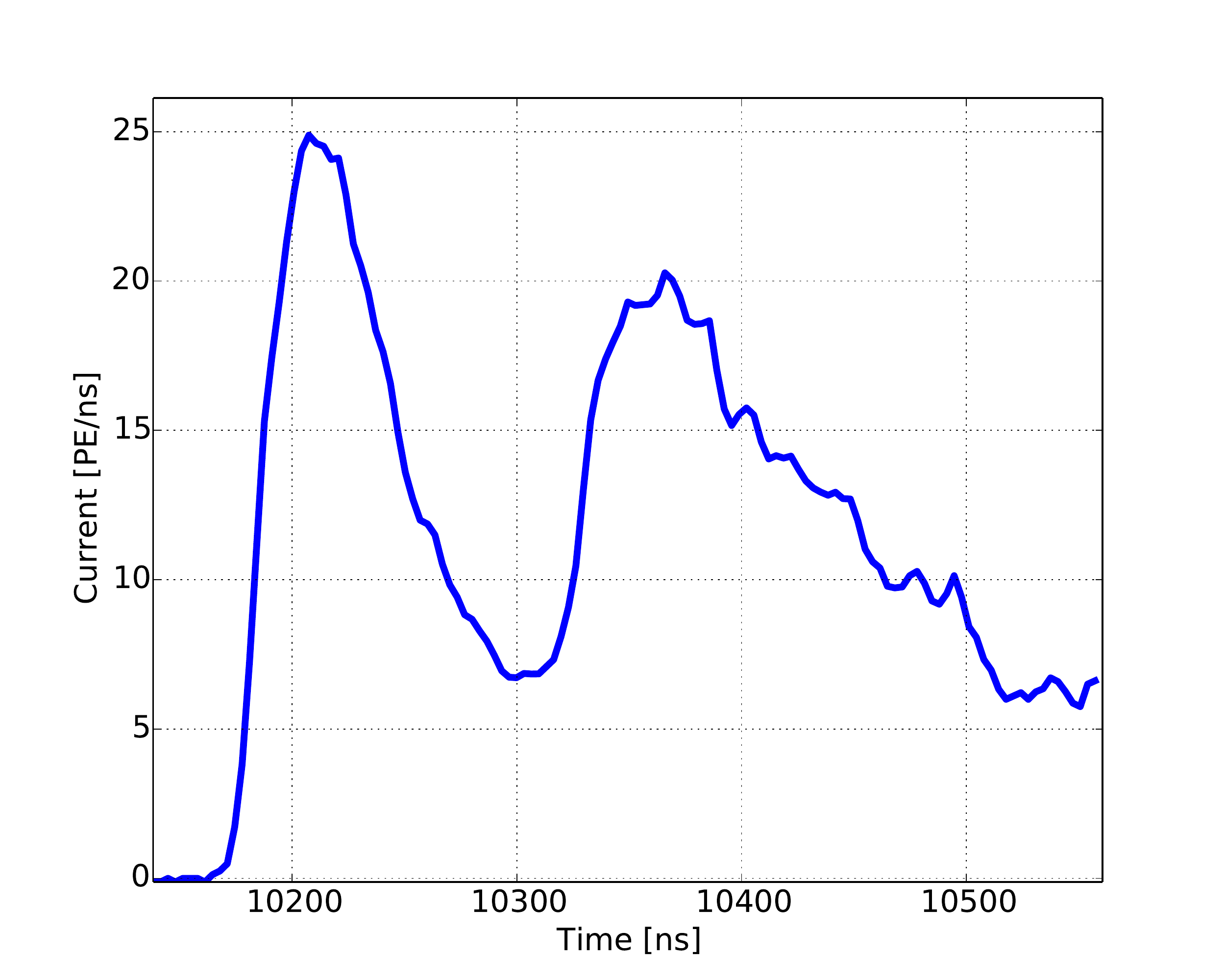} }}%
\caption{Left: A sketch of a tau neutrino undergoing CC interaction in
 IceCube. The figure is not drawn to scale. Right: A double pulse waveform
from a simulated $\nu_{\tau}$ CC event \cite{Aartsen:2015dlt}.}%
\label{fig:tau_dp}%
\end{figure*}

Neutrinos are difficult to detect, and they cannot be detected
directly. In IceCube, neutrino interactions are detected via the
Cherenkov radiation emitted by extremely relativistic secondary
particles which are produced by neutrinos interacting with the ice
nuclei. The identification of a neutrino interaction event
relies on the precise reconstruction of the event based on the timing
and charge information collected by the DOMs during an event readout. 
There are two major types of event topologies in IceCube. One is called
{\it track}, made by cosmic-ray induced atmospheric muons or
$\nu_{\mu}$-induced muons. These muons can penetrate long distances in
the ice at speeds close to the light speed in vacuum, emitting
Cherenkov light along their trajectory. The other is called {\it cascade}, or {\it
  shower}. Cascade events have high levels of degeneracy. They can be
made by charged current (CC) interactions of $\nu_e$, low energy
$\nu_{\tau}$, and neutral current (NC) interactions of all neutrino
flavors. During these interactions, hadrons or electrons are produced
at the interaction vertices, which subsequently interact with the ice and
produce cascades of secondary particles. The energy resolution for cascade and
track events above 100 TeV are $\sim$10\% and a factor of 2, respectively. 
While at these energies, the angular resolution for cascades and tracks are
15$^{\circ}$ and $<1^{\circ}$, respectively \cite{Aartsen:2013vja}.  
A third type of event topology called {\it double cascade}, or {\it
  double bang}, can be made by high energy $\nu_{\tau}$ CC
interactions \cite{learned1995detecting}. These types of events have not
been observed. A high energy $\nu_{\tau}$ undergoing CC
interaction in the ice produces a first cascade, and the outgoing
$\tau$ lepton\footnote{IceCube cannot discriminate neutrino and
  anti-neutrinos, so $\tau^{-}$ and $\tau^{+}$ will behave the
  same.} decays subsequently into hadrons or electrons with a total branching
ratio of $\sim$83\%\footnote{The $\tau$ lepton decays muonically 17\%
  of the time, producing an abruptly brightened track.},
producing a second cascade (left panel of
Fig.~\ref{fig:tau_dp}). The separation between the two cascades scales
as ($\frac{E_{\tau}}{1\,\mathrm{PeV}} \cdot 50$~m), with $E_{\tau}$
being the energy of the tau lepton. At energies below PeV, the double
cascade signature is difficult to distinguish from a single cascade,
due to the sparse spacing of DOMs. However, the double pulse process from
the $\nu_{\tau}$ CC interaction could manifest itself in the photon sensors as
double pulse waveforms. The right panel of
Fig.~\ref{fig:tau_dp} shows one example of such double pulse waveforms.

\section{A search for astrophysical tau neutrinos}

\subsection{Double pulse algorithm}

\begin{figure*} [t]
\centering
 \subfloat{{\includegraphics[width=0.47\textwidth]{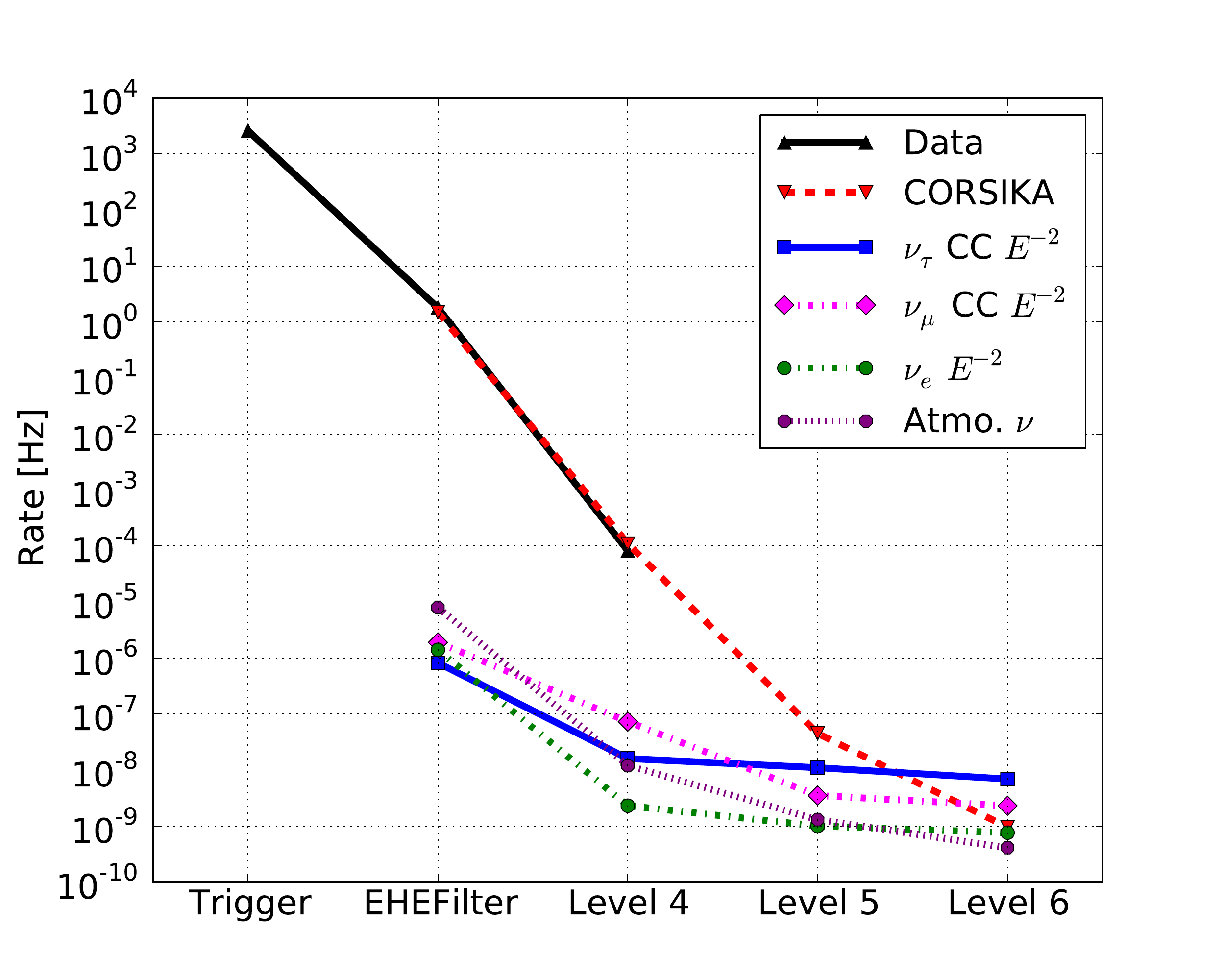}}}%
    \qquad
    \subfloat{{\includegraphics[width=0.47\textwidth]{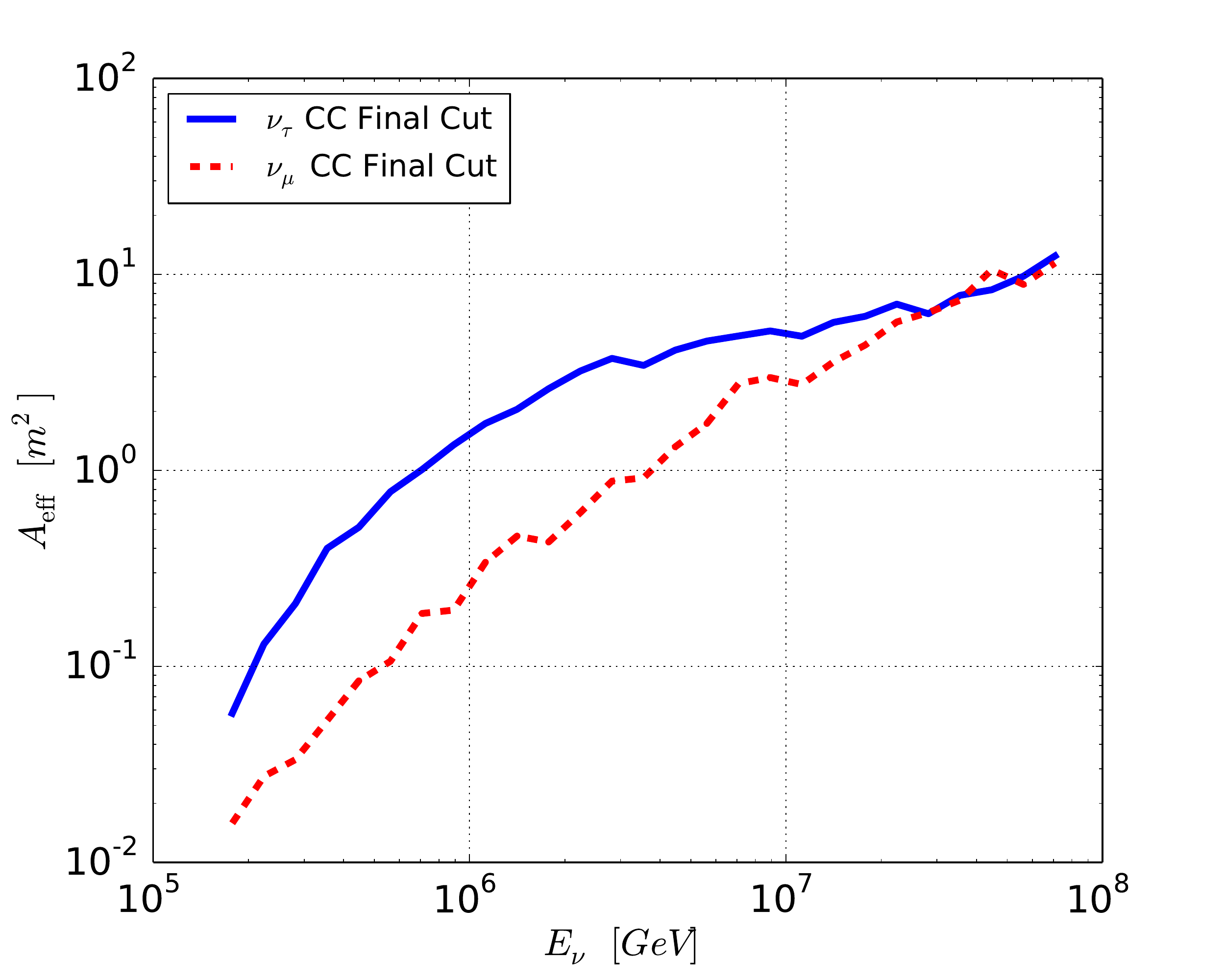} }}%
\caption{Left: Event passing rates at various cut levels. Right:
  Effective areas for $\nu_{\tau}$ CC and $\nu_{\mu}$ CC events
at final cut level \cite{Aartsen:2015dlt}.}%
\label{fig:dp_cuteff_aeff}%
\end{figure*}

The double pulse algorithm (DPA) is designed to identify bright
waveforms with double peak signatures that are consistent with
$\nu_{\tau}$ CC interactions, while rejecting bumpy waveforms caused
by late scattered photons from single energetic cascades (NC events of all
flavor and $\nu_e$ CC events). Below 1 PeV, it is extremely difficult to separate double cascades from single
cascades using event topologies. So it is of essential importance to bring the single cascade
backgrounds under control at the waveform level. 
A double pulse waveform is defined by a rising edge, followed
by a trailing edge and then followed by a second rising edge. 
A second trailing edge is not required as it usually runs outside of
the ATWD time window. The DPA uses 7 configurable parameters based on the first
derivatives of an ATWD waveform to characterize the rising and
trailing edges \cite{Aartsen:2015dlt, icrc2013:taudp}. 
The flow of the DPA is summarized as follows.
Since the $\nu_{\tau}$ CC events that could produce resolvable double pulse waveforms are usually close
  to a DOM, the waveforms are bright. Therefore, in order to increase
  computational efficiency, the DPA is only run on waveforms with an
  integrated charge greater than 432 photoelectrons (PE).
A sliding time window with bin size of 3.3 ns is employed to
determine the beginning of a waveform. The first rising edge is
considered starting if there is a monotonic increase
  in 6 bins, a duration of 3.3$\times$6=19.2~ns. 
Once the starting of the first rising edge is identified, the
  waveform is divided into segments with 4 ATWD bins (3.3$\times$4=13.2~ns), and the first derivatives are computed in each segment. 
The time segments of the first derivatives above/below zero, which
represent the duration of rising/trailing
  edges, are set to be at least 2, 2, 3 for the first rising, the
  first trailing and the second rising edges, respectively. 
The integrals of the continuous positive and negative first derivative
  values, which are representations of the steepness of the rising and
 trailing edges, are set to be at least 23 PE, 39 PE and 42 PE for the first rising, the first trailing and the second rising edges, respectively.
The DPA parameters were optimized
using a variety of IceCube event waveforms including neutrino Monte Carlos
of all flavors, simulated atmospheric muons, and {\it in situ}
calibration lasers.  

\subsection{Event selection and results}

\begin{figure*} [t]
\centering
 \subfloat{{\includegraphics[width=0.47\textwidth]{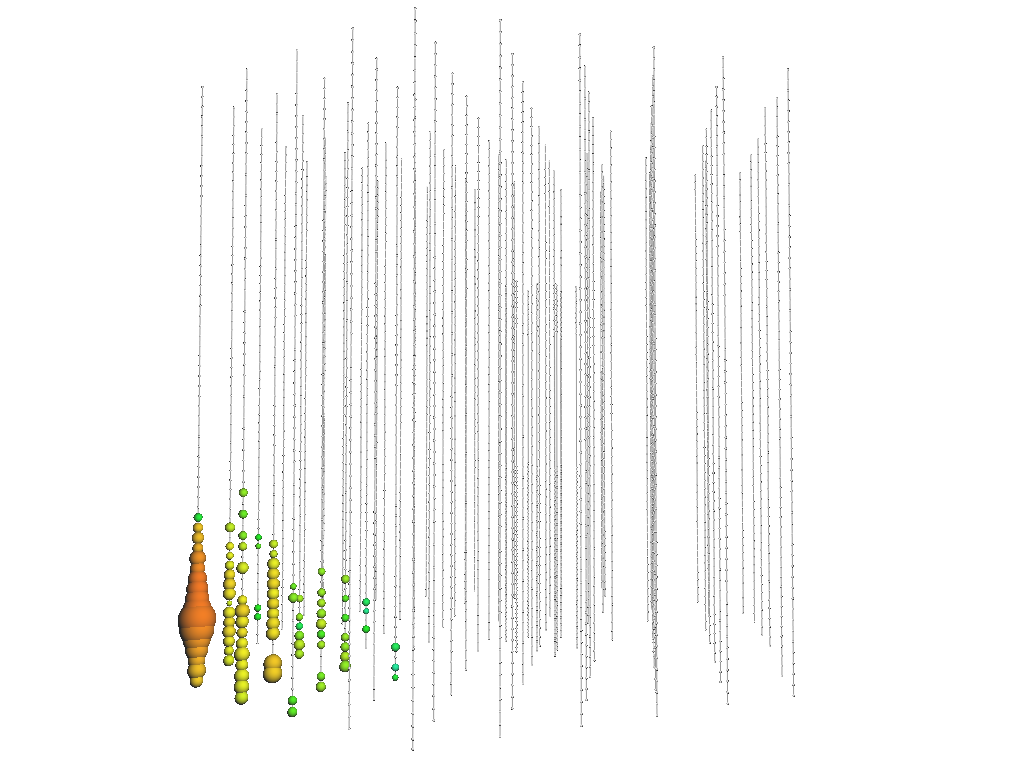}}}%
    \qquad
    \subfloat{{\includegraphics[width=0.47\textwidth]{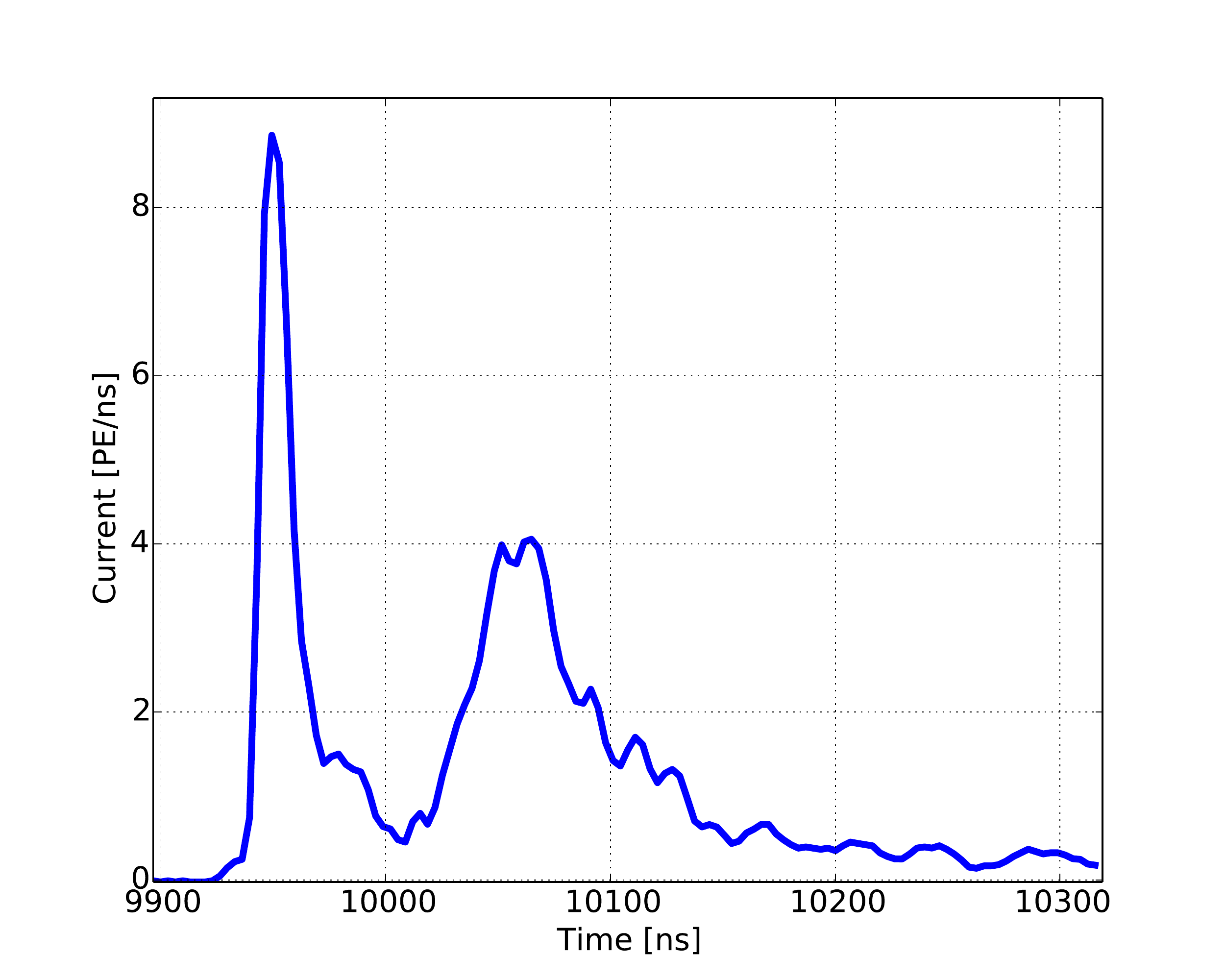} }}%
\caption{Left: A corner clipper event with double pulse waveforms at
  Level 5. Right: The corresponding double pulse waveform, which was from the
  brightest hit DOM of this event \cite{Aartsen:2015dlt}. }%
\label{fig:level5}%
\end{figure*}

The event selection criteria were developed using Monte Carlo samples of neutrinos,
atmospheric muons and 10\% of data (called the {\it burn sample}). To avoid introducing artificial
bias, the remaining 90\% of data were kept blind until the
selection methods were well tested and finalized. 
Data samples analyzed were collected between May 13, 2011 and May 6,
2014. Exclusions of the 10\% burn sample, data collection with only partial
detector operation or calibration runs when light sources were in use, resulted in a total of
914.1 days of livetime. 

\begin{figure}
\begin{floatrow}
\floatsetup[table]{capposition=top}
\capbtabbox{%
\centering
{%
 \RawCaption{\caption{Predicted event rates from all sources at final geometrical
  containment cut. The astrophysical $\nu_{\tau}$ CC double
  pulse events dominate at final cut, with the leading background
 being astrophysical $\nu_{\mu}$ CC double pulse events. Errors are
 statistical only. 
   }
  \label{table:evt_rate}}%
}
  \begin{tabular}[!T]{|c|c|} \hline
  Data samples & Events in 914.1 days \\ \hline
  Astrophysical $\nu_{\tau}$ CC & $(5.4 \pm 0.1) \cdot 10^{-1}$ \\
  Astrophysical $\nu_{\mu}$ CC & $(1.8 \pm 0.1) \cdot 10^{-1}$  \\
  Astrophysical $\nu_{e}$ CC & $(6.0 \pm 1.7) \cdot 10^{-2}$  \\
  Atmospheric  $\nu$ & $(3.2 \pm 1.4) \cdot 10^{-2}$  \\
 Atmospheric muons & $(7.5 \pm 5.8) \cdot 10^{-2}$  \\ \hline
  \end{tabular}
}

\ffigbox{%
 \includegraphics[width=0.5\textwidth]{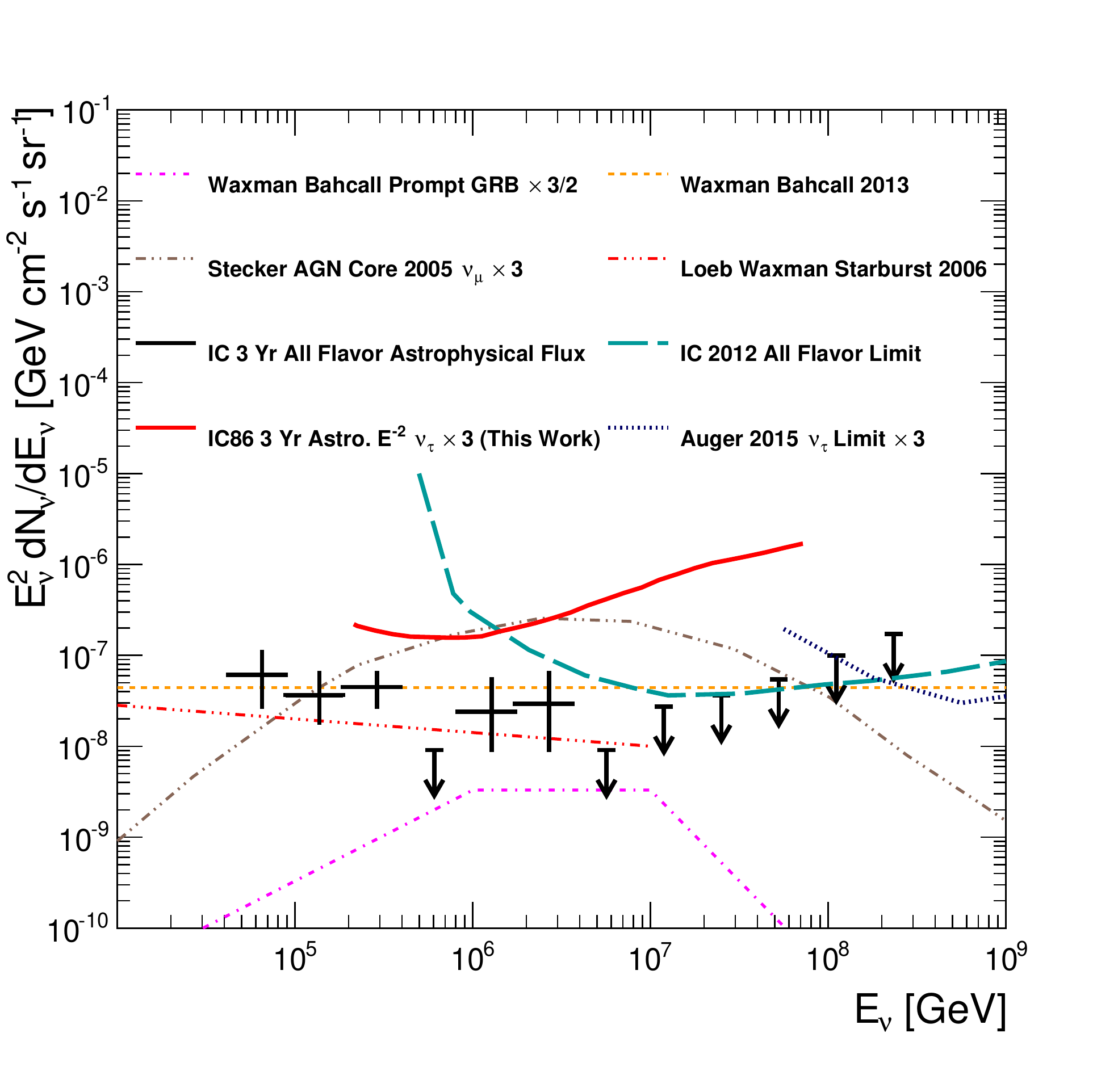}
}{%
  \caption{Differential upper limit on astrophysical $\nu_{\tau}$ flux at 90\% C.L
    (red solid line) \cite{Aartsen:2015dlt}.}%
  \label{fig:taudp_3yr}
}
\end{floatrow}
\end{figure}

The event selection began with the IceCube Extremely High Energy (EHE) filter, which requires an event has
at least 1000 PE. Event rates after this filter are $\sim$ 1 Hz. The
subsequent event selection process took three stages, named Level 4,
Level 5 and Level 6. At Level 4: an additional event-wise charge cut was placed at
$>$2000~PE. Then the DPA was applied and events with at least one
waveform passing were kept. At the Level 5 stage, the dominant
backgrounds are energetic atmospheric muons which lose energy
stochastically with two coincident energy losses making double pulse
waveforms. High energy $\nu_{\mu}$ CC events could also
make double pulse waveforms with the first pulse coming from the CC
vertex, and the second pulse from the coincident stochastic loss of
the outgoing energetic muon. A reduced likelihood ratio cut of L$_{\mathrm{R}}$
= (L$_{\mathrm{cascade}}$/L$_{\mathrm{track}}$) $<$ 0 was employed to separate
the cascade-like $\nu_{\tau}$ CC events from the track-like
atmospheric muon and $\nu_{\mu}$ CC double pulse
events. L$_{\mathrm{track}}$ and L$_{\mathrm{cascade}}$ are likelihood
values based on an infinite track and a point-like cascade hypothesis,
respectively. The first hit in an event was also required to be at
least 40~m below the top layer of the detector to further veto
downgoing atmospheric muons. At this level, the dominant background is
still atmospheric muons, with a total of 3.5$\pm$3.4 expected in 914.1
days. At the Level 6 stage: events surviving up to this stage are cascade-like and with at least one
double pulse waveforms. Particularly, the surviving cascade-like atmospheric
muons are the ones starting outside and clipping the corners of the
detector (called {\it corner clippers}). To eliminate these corner
clippers, a geometrical containment cut was cast which
required the reconstructed starting vetex of an event to be within
the detector's instrumented volume and some distance away from the
edge surface of the detector. The containment cut reduced the
atmospheric muon background significantly, yielding a total of $0.54$
signal $\nu_{\tau}$ double pulse events and $0.35$ total background double
pulse events in 914.1 days. The expected event rates are summarized in
Table~\ref{table:evt_rate}. 

Zero events were found upon unblinding the remaining 90\% data,
consistent with expectation. Based on these zero findings, for the
first time, a differential upper limit on the
astrophysical tau neutrino flux is set near the PeV energy region, as
shown in Fig.~\ref{fig:taudp_3yr}. The energy range encompassing the
middle 90\% of signal events is from 214 TeV to 72 PeV. 
At Level 5, three corner clipper events were
found, each of which has only one double pulse waveform. This finding
matches the Monte Carlo prediction of 3.5$\pm$3.4 in 914.1
days, indicating robustness of the analysis methods. 



\newfloatcommand{capbtabbox}{table}[][]

\section{Conclusion and outlook}

Searching for astrophysical tau neutrinos in the IceCube waveforms is
shown to be robust, with Monte Carlo predictions matching experimental
results.
Zero events were found at the final cut, consistent with the expectation of 0.5 signal events in
three years of IceCube data. More sophisticated methods including
machine learning algorithms are under
development to identify more ambiguous $\nu_{\tau}$ double pulse waveforms from the
enormous number of waveforms. Analyses to search for resolvable double
bang event topologies via precise reconstructions of event
vertices are well under way. With a suite of promising techniques, and
continuous accumulation of data, IceCube is getting closer than ever
to discovering astrophysical tau neutrinos.


\bibliographystyle{JHEP}
\bibliography{ichep2016}

\end{document}